\documentclass[journal,comsoc]{IEEEtran}
\usepackage[T1]{fontenc}
\usepackage{amsmath,amsthm}
\usepackage[cmintegrals]{newtxmath}
\usepackage{cite}
\usepackage[utf8]{inputenc}
\usepackage[english]{babel}
\usepackage{graphicx}
\usepackage{subcaption}
\usepackage{algorithm,algorithmic}
\usepackage{caption}
\usepackage{bm}
\usepackage{enumitem}

\newtheorem{theorem}{Theorem}[section]

\newtheorem{lemma}[theorem]{Lemma}

\floatname{Procedure}{algorithm}

\begin{document}
\title{A Robust Algorithm for Sniffing BLE Long-Lived Connections in Real-time}
\author{\IEEEauthorblockN{Sopan Sarkar, Jianqing Liu, Emil Jovanov} \\
\IEEEauthorblockA{Department of Electrical and Computer Engineering, \\
University of Alabama in Huntsville, Huntsville, Alabama 35758 USA \\
Email: \{sopan.sarkar, jianqing.liu, emil.jovanov\}@uah.edu}}

\maketitle

\begin{abstract}
Bluetooth Low Energy (BLE) has become an intrinsic wireless technology for the Internet of Things (IoT). With the proliferation of BLE-embedded IoT devices, it is important to study the security and privacy implications of BLE. The forefront attack to BLE devices is the wireless sniffing attack, which would lead to more detrimental threats like jamming, encryption cracking or system penetration. Existing sniffing attacks are based on the correct detection of BLE connection initiation state, but they become ineffective for BLE long-lived connections. In this paper, we focus on the adversary setting with a low-cost single radio and develop a suite of real-time algorithms to determine the key parameters necessary to follow and sniff a BLE connection in the connected state. We implement our algorithms in the open source platform \textemdash Ubertooth One and evaluate its performance in terms of sniffing overhead and accuracy. By comparing with state-of-the-art schemes, experimental results show that our sniffer achieves much higher sniffing accuracy (over 80\%) and better stability to BLE operational dynamics.
\end{abstract}


\IEEEpeerreviewmaketitle

\section{Introduction}
Since its inception, Bluetooth has evolved continuously, creating innovative ways for things to connect. BLE was first introduced in version 4.0 of the Bluetooth Standard \cite{ble2019spec}. It differs from classic Bluetooth with respect to modulation mode and link layer packet format, and was developed mainly targeting low power embedded devices. 

Due to its ubiquitous presence, reliability \cite{blereliability}, significant reduction in energy consumption and mesh networking capability  \cite{mesh}, BLE has become the dominant wireless connectivity standard for IoT. In 2018 alone, 4.7 billion BLE devices were shipped which is estimated to double by 2020 \cite{blestat}. Such a wide adoption rate unfortunately raises security and privacy alarm. For instance, BLE fitness trackers could become the target by adversaries for location tracking \cite{zhou2014security}; cellphone could be hijacked by injecting malformed packets via BLE interface \cite{seri2017blueborne}. Thus, it becomes imperative to assess the security and privacy implications of BLE devices.

Bluetooth specification defines several security mechanisms for authentication and link layer security, but it is often seen that developers ignore these security measures to trade for design simplicity and to obtain fast market access. There is some well-documented case of attacks on BLE devices that comprise of intervening BLE connections, stealing private information \cite{blethreats1, blethreats2, blethreats3} and jamming \cite{jamming}. These active attacks, although detrimental, can be easily detected by intrusion detection systems \cite{bleguard, misusedetect}.

An interesting but fully unexplored attack is the BLE sniffing, which is stealthy and intends to compromise user privacy or cause further system penetration. However, it is very difficult to sniff and follow a BLE connection because BLE employs the adaptive frequency hopping (AFH) spread spectrum scheme where it rapidly switches the carrier frequency in a pseudo-random manner. The parameters for determining the AFH pattern are only exchanged during the establishment of connection in the initiating state. Moreover, to avoid power interference in ISM band, AFH frequently modifies the hopping pattern which makes it even more difficult to sniff. 

There are several possible ways to sniff a BLE connection. One way is to use wide-band sniffers that can sniff multiple channels at a time. This may not be prohibitively expensive to less-capable adversaries (e.g., \$25K for a Bluetooth protocol analyzer\cite{fte}). There also exist some low-cost, open source but narrow-band sniffers \cite{adafruit, cypressdongle, ti} which either requires to establish a connection with the BLE victim device or observe the establishment of a BLE connection to extract the AFH parameters. If BLE victim devices are already in connected state, one way is to jam the connection and force it to re-pair and thus observe the AFH parameters \cite{mike}. Such an attack is known as re-pair attack and as stated earlier can be easily detected and avoided by intrusion detection systems. Moreover, such jamming is ineffective as the attacker has no idea of the BLE hopping pattern (unless it jams all BLE channels). In addition to that, due to advertisements on 3 channels, the probability of detecting the connection parameters decreases to 33.3\%. This is even greatly reduced in the presence of co-channel interference.

In this paper, we investigate the stealthy sniffing of a BLE connection that has been established for a while. We present the design, implementation, and evaluation of our algorithm that utilizes a single-channel and low-cost radio device for determining the BLE AFH pattern. The algorithm basically observes a single channel at a time and determines the AFH parameters namely connection interval ${c_{int}}$, hop increment ${h_{inc}}$ and the channel map ${c_{map}}$ in real-time. Next, we implement our design in the open source platform \textemdash Ubertooth One \cite{uber} and evaluate its performance.

The rest of the paper is organized as follows. Section II introduces the most recent literatures on this topic. Section III provides background on BLE communication focusing on the link layer specification. The proposed algorithms for determining the BLE AFH parameters are described in Section IV. We describe our experimental methodology and evaluate its performance in Section V. Finally, Section VI concludes the paper.

\section{Related Works}
Bluetooth has enjoyed high penetration rate in smartphones and consumer electronics in recent years. This has sparked interest among researchers to look into the privacy and security implications of Bluetooth. BlueEar \cite{blueear} is a (classic) Bluetooth traffic sniffer developed based on a dual-radio architecture. One of the radios called scout hops through all the channels to monitor interference and target's packet transmission, which is then used by the other radio called the snooper to hop and follow the target. The system uses probabilistic clock matching algorithm to determine basic hopping pattern, and statistical model with a lightweight machine learning algorithm to predict adaptive hopping behavior. Their method shows a packet capture rate of 90\% in a real-world environment.

With the recent growth of interest in IoT devices, the focus of deployment quickly shifted towards BLE. BLE specification presents security mechanisms for device authentication, data encryption and redundancy checks at the link layer. However, recent studies have shown the possibility to bypass such security measures. Das \emph{et al.} in \cite{fitnesstracker} demonstrate how BLE traffic of fitness trackers can be correlated with the intensity of user's activity and thus be used to identify and track him/her in the crowd. In Das's experiment, they use off-the-shelf, expensive and proprietary packet sniffers to observe control packets from all three advertisement channels.

Aforementioned sniffing attacks are based on multi-radio platforms which could be prohibitively expensive for less-capable adversaries. Currently, one popular platform \textemdash Ubertooth One is an open-source, single-radio and cheap Bluetooth sniffer that was developed by Ryan \emph{et al.} \cite{mike}. This is a powerful sniffer that is relied on observing advertisement packets and looking for AFH parameters to follow.  As stated earlier it is difficult to sniff a BLE connection after the connection has been already established. In \cite{mike}, Ryan \emph{et al.} imply how AFH parameters can be extracted from a long-lived BLE connection via jamming. Moreover, Ryan \emph{et al.} also demonstrate a proof-of-concept of key re-negotiation and man-in-the-middle (MitM) attack to BLE devices.

Bluetooth Low Energy Multi (BLE-Multi) \cite{Blemulti} is a firmware extension to Ubertooth One which enables tracking of multiple simultaneous long-lived connections. It uses transmission of empty packets to determine the anchor point of each connection event and connection timing. Moreover, it achieves multi-connection sniffing by opportunistically switching between connections when they move from active to sleep mode. Similar researches include the work by Cheng \emph{et al.} \cite{tsch} which demonstrate an attack to IEEE 802.15.4e (Zigbee) where they crack the TSCH channel hopping algorithm based on a wide-band receiver to monitor all the channels.

Our work is similar to that of Ryan's, but we consider a more realistic environment and tackles all the complexities associated with it. One of the important assumptions made by Ryan during his demonstration is that BLE connections use all 37 data channels, but in reality this is not the case due to AFH. Therefore, the algorithms provided by Ryan perform poorly in the practical environment. In this work, we consider a realistic situation where BLE adopts adaptive frequency hopping and the connection has been established for a while. Our algorithm is designed to extract the AFH parameters on-the-fly. Though our algorithm is deployed on the Ubertooth One platform, it is platform independent and can be deployed on any system.

\section{BLE Preliminaries}
BLE operates in the 2.4 GHz ISM band at frequency 2408 – 2483.5 MHz. The whole spectrum is divided into 40 channels with 2 MHz spacing and indexed from 0 to 39. Among these channels, three channels (channel 37, 38 and 39) are used for advertising and the rest 37 channels are used for data communications. In this section, we will provide an overview of the BLE advertising and data communication process.

\begin{figure}[hbt!]
  \begin{center}
  \includegraphics[width=3.5in]{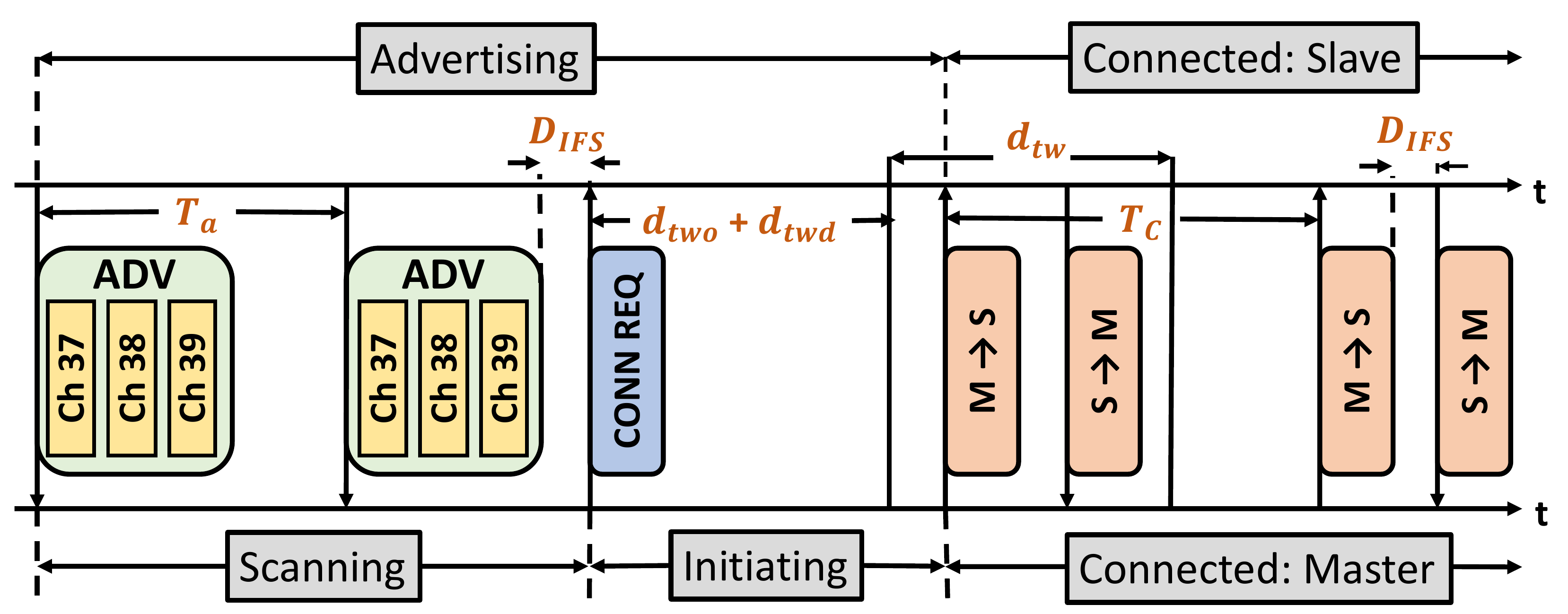}
  \end{center}
  \begin{center}
   \parbox{8cm}{\caption{BLE Link Layer states with connection establishment procedure and data flow.}\label{ble}}
  \end{center}
\end{figure}

Fig. \ref{ble} gives an overview of the BLE Link Layer states along with the communication process. A device in the standby state does not transmit or receive packets. When a device is in advertising state, it continuously broadcasts advertisement packets in the advertising channels. The BLE specification recommends to use all three advertising channels, but any manufacturer can use less than three channels for energy efficiency at the cost of reliability. Advertisements (a.k.a., beacons) are sent periodically at ${T_a}$ intervals, which consist of a fixed delay of 20ms – 10.24s and a random delay of 0 – 10ms. The random delay ensures the reduction of collision of advertisement packets from multiple devices. Both BLE advertisement and data packets have the same format that contains a preamble, access address (AA), packet data unit (PDU) and cyclic redundancy check (CRC). The AA for the advertisement packets is fixed (0x8E89BED6) and for the data packets, it varies for each connection. The time interval between two consecutive packets on the same channel is 150us and is denoted by ${D_{IFS}}$.

A device in the scanning state continuously listens to advertisement packets. It may query further information from the advertiser through a scan request (SCAN\_REQ) packet, to which the advertiser responds with a scan response (SCAN\_RSP) message.

When a device wants to establish a connection, it goes to the initiating state and sends a connection request packet (CONNECT\_REQ) to the advertiser that contains the required parameters to establish the connection. When the request is accepted by the advertiser, both the devices go to the connected state. The first regular connection event is scheduled after ${d_{two}}$ + ${d_{twd}}$ time from the CONNECT\_REQ and within the ${d_{tw}}$ interval. In this state, the advertiser takes the slave role whereas the initiator as the master.
In the connected state, BLE devices hop to a new channel every ${c_{int}}$ time interval to exchange packets following the channel selection algorithm as specified in the BLE specification. The ${c_{int}}$ is selected between 7.5ms to 4s through negotiation between the master and slave. The BLE channel selection algorithm is used to select the next channel to be hopped to. It takes in three parameters: ${c_{map}}$, ${h_{inc}}$ and last unmapped channel ${luc}$. The ${c_{map}}$ is a bit sequence of 40 bits corresponding to the 40 BLE channels. It classifies the channels as Used and Unused based on the master's channel assessment algorithm. If a channel is Used, a bit value of ‘1’ is assigned to its position value and vice versa. For any BLE connection, the minimum number of channels to be used is two. The ${h_{inc}}$ is a number randomly selected between 5 and 16 which specifies the number of channels to be skipped on each hop. It is to be noted that for a particular connection the  ${c_{int}}$ and ${h_{inc}}$ remains constant. The ${c_{map}}$ can be updated through a channel map update request packet (LL\_CHANNEL\_MAP\_REQ). The BLE channel selection algorithm is illustrated in Fig. \ref{ch_al}. 
\begin{figure}[hbt!]
  \begin{center}
  \includegraphics[width=3.5in]{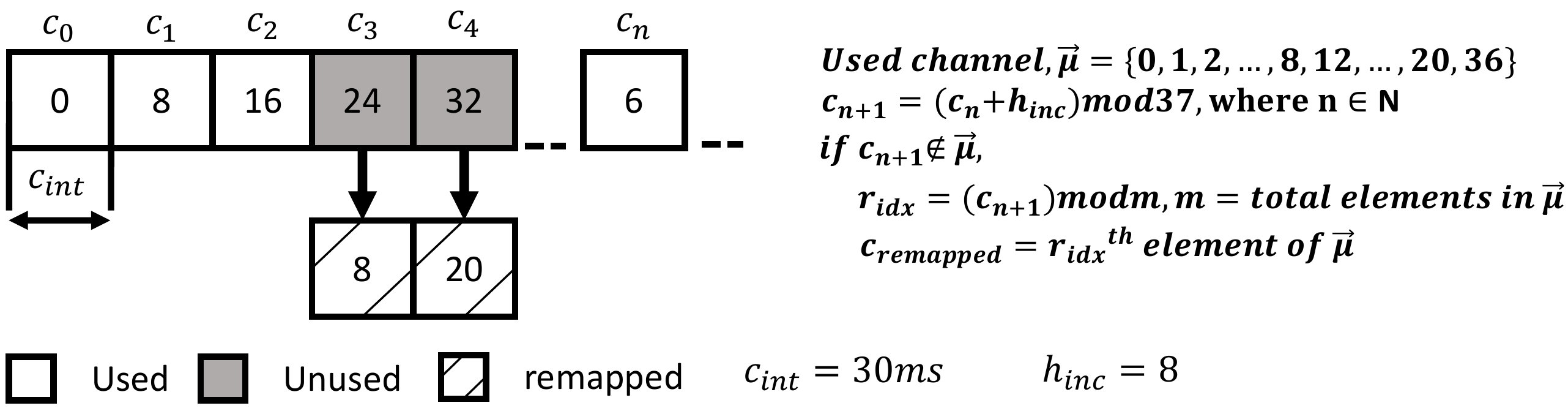}
  \end{center}
  \begin{center}
   \parbox{8cm}{\caption{BLE channel selection algorithm.}\label{ch_al}}
  \end{center}
\end{figure}

\section{Real-time BLE Sniffing Algorithm}
\subsection{System Model}
We consider an indoor environment where BLE devices have limited mobility, if not fully static. Most BLE devices are in the connected state to exchange data while some of them could be in advertisement state to broadcast beacon signals. For the former ones, AFH is employed to select carrier frequency to transmit data whereas the latter ones follow fixed frequency hopping pattern (channel 37 $\rightarrow$ 38 $\rightarrow$ 39 $\rightarrow$ 37) to broadcast messages. All BLE devices (v4.0 or beyond) apply 128-bit AES for payload encryption but the packet header is in plaintext.

\subsection{Attack Model}
Suppose an adversary, Eve, stays in close proximity to victim BLE devices (e.g., in the same building). Eve has limited resources in terms of accessible equipments and computation capability. Specifically, we assume Eve carries a single-radio sniffer with a simple protocol analyzer like Wireshark. The adversarial objective is to follow a target BLE device's connection and sniff its airborne packets, which could lead to location tracking. Although the attack is passive and stealthy, Eve may latter launch jamming attacks to disconnect the connection or inject malformed packets to take over victim BLE devices. For those in the advertisement state, sniffing is trivial. The challenge to Eve is to crack the target's AFH pattern when it is in the connected state, which is the focus of this work.

\subsection{Real-time BLE Sniffing Algorithms}
Here is the overview of our idea. First, Eve stays on a BLE channel and filters out the target via its AA or MAC address in the plaintext header. Next, Eve attempts to crack AFH pattern by determining ${c_{int}}$, ${h_{inc}}$ and ${c_{map}}$ one after another. The nature of AFH parameters is that ${c_{int}}$ and ${h_{inc}}$ are constant for an established connection while ${c_{map}}$ varies over time depending on the power interference level. As a result, the update of ${c_{map}}$ would totally change the hopping pattern and thus making it difficult to sniff a BLE connection. In what follows, we discuss how to address this challenge and derive corresponding AFH parameters.

\subsubsection{Determining ${c_{int}}$}
The hopping pattern of BLE connection is repetitive for a given channel map ${c_{map}}$. We leverage this property to extract the connection interval ${c_{int}}$ during a grace period when ${c_{map}}$ remains unchanged . In this section, we first prove that BLE hopping pattern is repetitive for a given channel map and then describe the algorithm for the derivation of connection interval.

\begin{lemma}
BLE channel hopping has a repetition period of 37 for a given channel map.
\end{lemma}
\begin{proof}
According to the Linear Congruential Generator algorithm \cite{lcg}, the sequence of pseudo-random values is given by,
\begin{equation}\label{ln_generator}
{x_{n+1}} = \left({ax_n} + c \right) \, \text{mod} \, m
\end{equation}
when, ${c}$ $\ne$ 0, ${x_n}$ will have a period of ${m}$ if and only if
\begin{enumerate}[label=(\roman*)]
	\item ${m}$ and ${c}$ are relatively prime
	\item ${a-1}$ is divisible by all prime factors of ${m}$ and
	\item ${a-1}$ is divisible by 4 if ${m}$ is divisible by 4
\end{enumerate}
In BLE AFH, the channel selection algorithm is given by,
\begin{equation}\label{ch_al}
{c_{unmapped}} = {({luc} +  {h_{inc}}) \, \text{mod} \, 37},
\end{equation}
It has the same form as Eq. (\ref{ln_generator}) for ${a}$ = 1. Moreover, given 37 is a prime number, it satisfies the conditions (i) and (ii). Thus, we claim that the AFH pattern in BLE communication has a period of m = 37.
\end{proof}
\begin{figure}[hbt!]
  \begin{center}
  \includegraphics[width=3in]{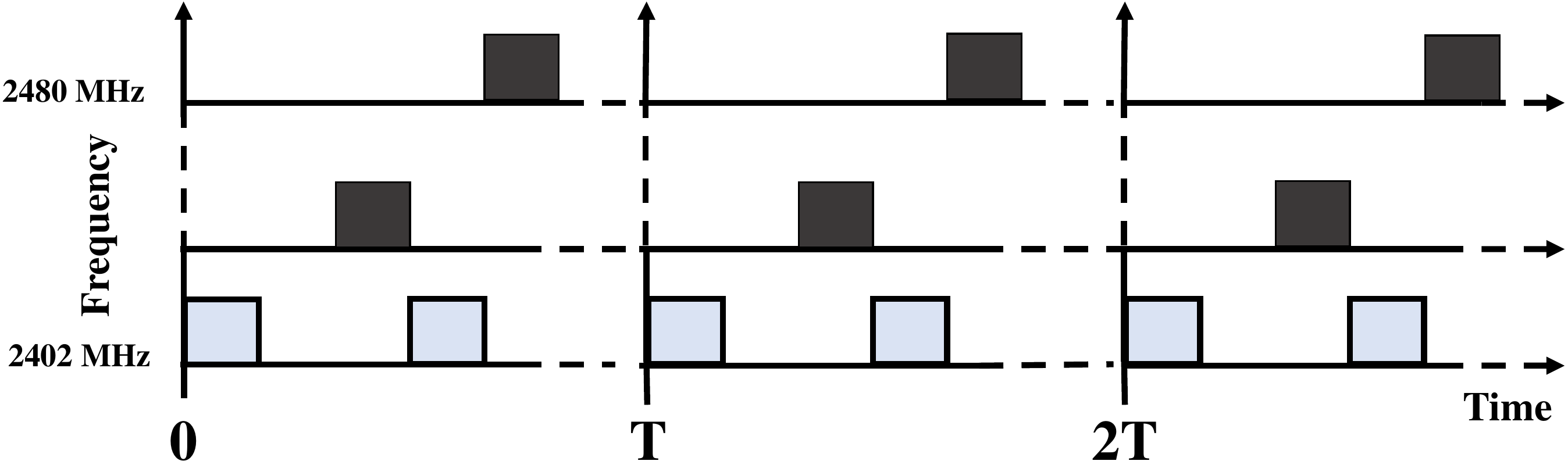}
  \end{center}
  \begin{center}
   \parbox{8cm}{\caption{BLE frequency hopping diagram where the blue block represents the observed channel}\label{con_int}}
  \end{center}
\end{figure}
Fig. \ref{con_int} shows the repetitive property of a BLE connection. When there is no ${c_{map}}$ update, the hopping pattern repeat itself every ${T}$ intervals, where ${T}$ = 37*${c_{int}}$.
For determining connection interval, we set the sniffer to stay at single frequency channel (e.g., channel 0) and record the time difference $\Delta t$ between adjacent packet receptions. We then compare and analyze the set of $\Delta t$' in real-time and find the repetitive pattern among them. For accuracy, we compare this pattern for a defined repetitions (e.g., 3). The sum of all $\Delta t$' in a single pattern is ${T}$ which accounts for 37 hops. Then, the connection interval is calculated as
\begin{equation}
{c_{int}} = \frac{1}{37}\sum \Delta {t}
\end{equation}

As shown in Fig. \ref{con_int}, it is important to note that a channel may appear multiple times within a 37-hop period (i.e., $T$) when the number of used channels in ${c_{map}}$ is less than 37. Among these appearances, only one channel is truly mapped while the rest are re-mapped.

\subsubsection{Determining ${h_{inc}}$}
Suppose at time ${t_1}$ we observe channel ${c_1}$ and then we jump to channel ${c_2}$ and observe it at time ${t_2}$. Then, the total number of hops from ${c_1}$ to ${c_2}$ is given by
\begin{equation}
h =  \frac{\Delta{{t_2}} - \Delta{{t_1}}}{c_{int}}.
\end{equation}
According to \cite{mike}, we can determine the ${h_{inc}}$ as follows:
\begin{equation}\label{eqn_hi}
{h_{inc}} = \frac{\left({c_2} - {c_1}\right)^{-1} }{h} \, \text{mod} \,37
\end{equation}

\begin{figure}[hbt!]
  \begin{center}
  \includegraphics[width=3in]{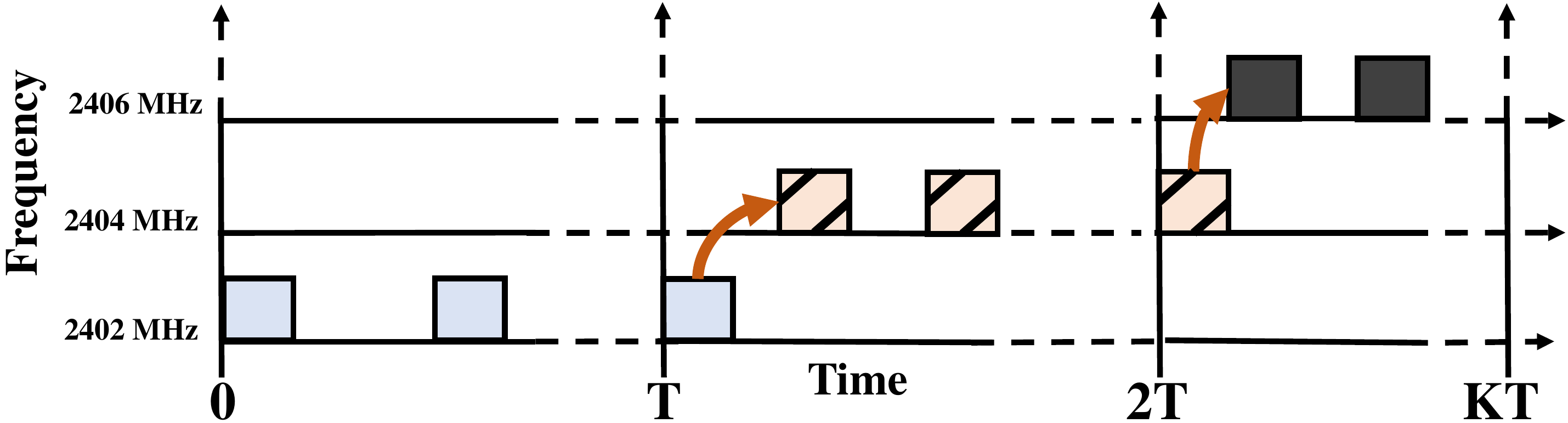}
  \end{center}
  \begin{center}
   \parbox{8cm}{\caption{Determining hop increment given multiple appearances of a channel within period $T$}\label{hop_inc}}
  \end{center}
\end{figure}

Here, the integer ${h_{inc}} \in [5,16]$. If the channel ${c_1}$ and ${c_2}$ appear only once within period $T$, they are truly mapped channels and we can determine ${h_{inc}}$ using Eq. (\ref{eqn_hi}). However, in reality, channel map ${c_{map}}$ generally contains less than 37 channels and each channel can appear multiple times within $T$, as the channel 0 (blue block) shown in Fig. \ref{con_int}. In this case, Eq. (\ref{eqn_hi}) would give a wrong ${h_{inc}}$ if any of channel ${c_1}$ and ${c_2}$ is remapped. 
In general, suppose there are ${n}$ appearances in ${c_1}$ and ${m}$ appearances in ${c_2}$ within each period $T$. We would have $n \times m$ hop increment values for any combination of these channels. It is mathematically infeasible to determine which ${h_{inc}}$ is the correct one, although we could narrow down the searching space by eliminating ${h_{inc}} \notin [5,16]$. In light of this, if observing two channels would not give a valid while unique ${h_{inc}}$, we observe a third channel ${c_3}$ which may also appear ${l}$ times within period $T$, as shown in Fig. \ref{hop_inc}. In so doing, we find the hop increment ${h_{m,l}}$ from channel ${c_2}$ and ${c_3}$. By intersecting two results, namely ${h_{n,m}}$ and ${h_{m,l}}$, the correct ${h_{inc}}$ would be identified. Although we may need to observe more channels in extreme cases, our empirical study shows ${h_{inc}}$ is found by observing at most three channels by applying a probabilistic rule that the mostly appeared (over 50\%) ${h_{inc}}$ is the correct one. In this case, the specific combination of ${c_1}$, ${c_2}$ and ${c_3}$ that gives the correct ${h_{inc}}$ are true channels.
To give a clear picture, the algorithm for determining ${h_{inc}}$ is stated below.
\begin{algorithm}
\caption{Algorithm for determining ${h_{inc}}$}
\begin{algorithmic}[3]
    \REQUIRE Initialize set $P$ containing observations of ${c_1}$
    \ENSURE  ${h_{inc}}$
\STATE hop to channel ${c_2}$ and observe for time $T$ and generate set $Q$ containing observations of ${c_2}$;
\STATE hop to channel ${c_3}$ and repeat prior step to generate set $R$;
\FOR {$n \in P$ and $m \in Q$}
	\STATE calculate ${h_{n,m}}$ = $\frac{\left({c_2} - {c_1}\right)^{-1} }{h} \, \text{mod} \,37$
	\IF {${h_{n,m}} \in [5,16]$ }
		\FOR {$l \in R$ }
		 	\STATE calculate ${h_{m,l}}$ = $\frac{\left({c_3} - {c_2}\right)^{-1} }{h} \, \text{mod} \,37$
			\IF {${h_{n,m}} = {h_{m,l}}$}
				\STATE store ${h_{n,m}}$ in set $J$
				\STATE save $<n, m, l>$ as true channels
			\ENDIF
		\ENDFOR
	\ENDIF
\ENDFOR
\end{algorithmic}
\end{algorithm}

\subsubsection{Determining ${c_{map}}$}
To get ${c_{map}}$, we let the sniffer mimic victim BLE devices' behavior that runs the same channel hopping algorithm as in Eq. (\ref{ch_al}) based on the derived ${c_{int}}$ and ${h_{inc}}$. In so doing, after 37 hops, the attacker shall know which channels are in ${c_{map}}$ and which are not. However, the starting channel $luc$ must be a truly mapped channel; otherwise, the attacker loses synchronization with the victim BLE connection. Therefore, after determining ${h_{inc}}$, we first move to a truly mapped channel $luc$ determined in the previous step. We then use Eq. (\ref{ch_al}) to calculate the next channel to be hopped and move to that channel earlier than the victim BLE devices (e.g., $\frac{c_{int}}{2}$ sooner). This ensures that we have enough time to observe the connection on that channel. We repeat the above process till completing all 37 hops. If we observe target AA in any channel we mark it as Used, otherwise we mark it as Unused. Upon completion, the set of Used channels is the channel map of the current BLE connection. The algorithm for determining the channel map is stated below.

\begin{algorithm}\label{map_algo}
\caption{Algorithm for determining of ${c_{map}}$}
\begin{algorithmic}[4]
    \REQUIRE Initialize $luc$ = observed truly mapped channel, ${h_{inc}}$
    \ENSURE ${c_{map}}$
\FOR {count = 1, count ${\leq}$ 37, count++}
	\STATE set channel = $(luc + {h_{inc}}) \, \text{mod} \, 37$
	\IF {packet containing target AA is observed}
		\STATE mark the channel as Used
		\STATE append it to ${c_{map}}$
	\ELSE
		\STATE set the channel as Unused
	\ENDIF
\ENDFOR
\end{algorithmic}
\end{algorithm}

Now we know all parameters required to follow a BLE connection. One important thing to note is that when channel map updates, we will not be able to immediately detect it as this information is encrypted in the payload. As a result, we may lose synchronization with the BLE connection and our sniffing performance will be hampered. To mitigate this problem, we keep track of the number of missed packets in a sliding window. If it exceeds a certain threshold, we assume that a channel map update has taken place and we re-run Alg. 2 to calculate the channel map. Recall that ${c_{int}}$ and ${h_{inc}}$ do not change during the BLE connection.
\section{Performance Evaluation}
In this section we evaluate the performance of our algorithm. We first describe the setup and evaluation metrics, and then discuss results in details.
\subsection{Experiment Setup}
\begin{figure}[hbt!]
  \begin{center}
  \includegraphics[width=2.3in]{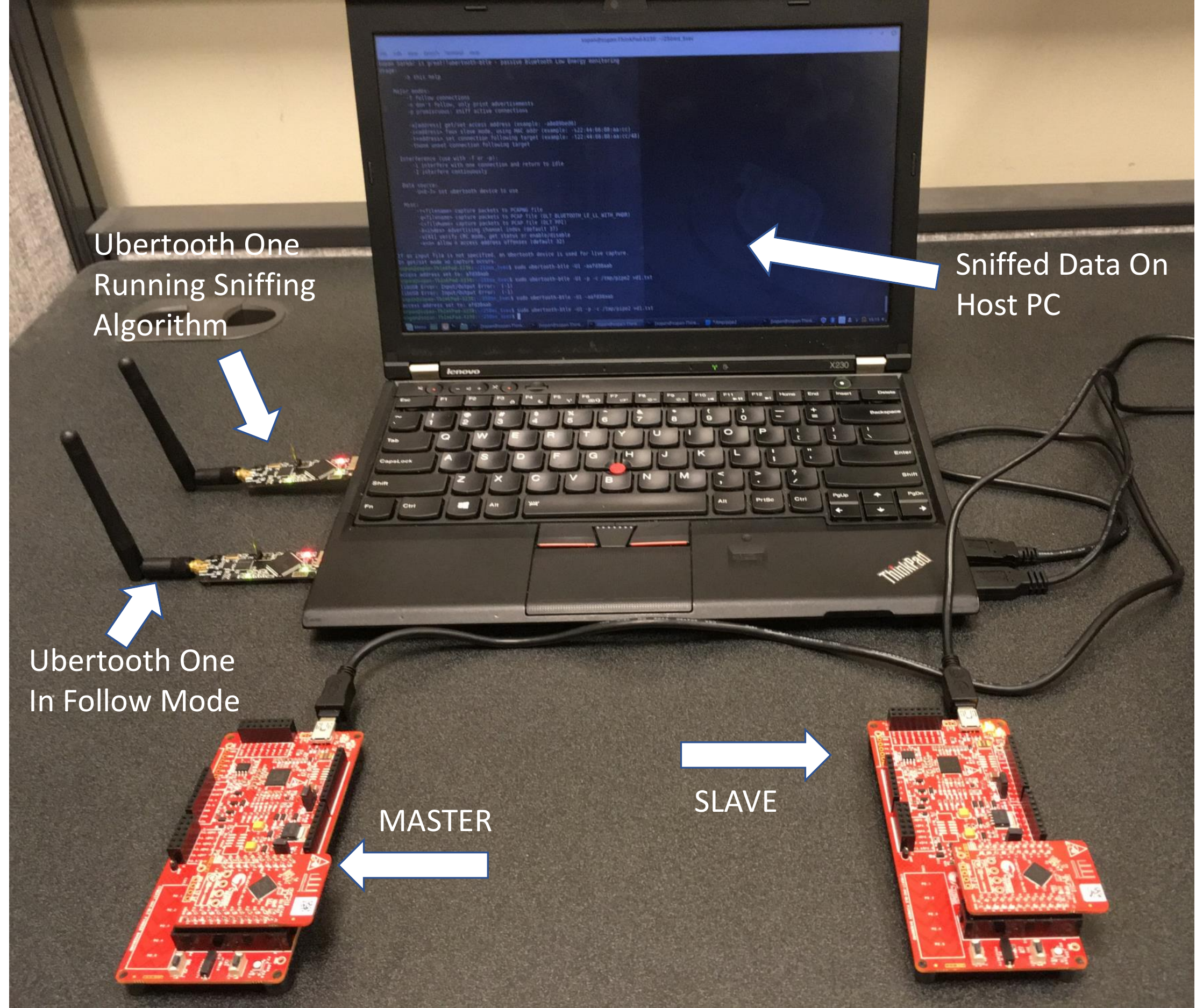}
  \end{center}
  \begin{center}
   \parbox{8cm}{\caption{Experiment setup in an indoor controlled environment}\label{exp_setup}}
  \end{center}
\end{figure}

We implement our algorithm in the open source platform Ubertooth One. To develop the firmware and program, we use the Ubertooth One with a Linux based PC. As shown in Fig. \ref{exp_setup}, our experiment platform consists of two Ubertooth One devices connected to a Linux PC. One of them runs the default Follow Mode (as benchmark) whereas the other one runs our developed algorithms. We use Cypress BLE platforms as test devices and setup one device as the master and another one as the slave. The main reason for choosing Cypress is that it provides a controlled environment that can freely change AFH parameters. This gives us enough flexibility to test our algorithms for different settings.

We carry out several experiments that were necessary to understand the BLE characteristics and evaluate the performance of our developed algorithms. Specifically, we perform the following experiments:
\begin{enumerate}
	\item We evaluate the accuracy and latency of our algorithms in extracting the AFH parameters. In this case, we set the ${c_{map}}$ to update every 15s, 30s and 60s respectively.
	\item Next, we compare the performance of our system in sniffing data packets with that of Ubertooth One in follow mode. The follow mode needs to observe BLE advertisements and look for a connection request packet from which it will extract all the required AFH parameters. After that, it goes to the data channel and starts to follow the connection. We then run our sniffer, which starts to follow the connection after extracting AFH parameters from data channels. Now, we send 100 data packets every 100ms in the first setting and 250ms in the second one.
\end{enumerate}
\begin{figure*}[hbt!]
\centering
\begin{minipage}[b]{.3\textwidth}
\includegraphics[width=2in]{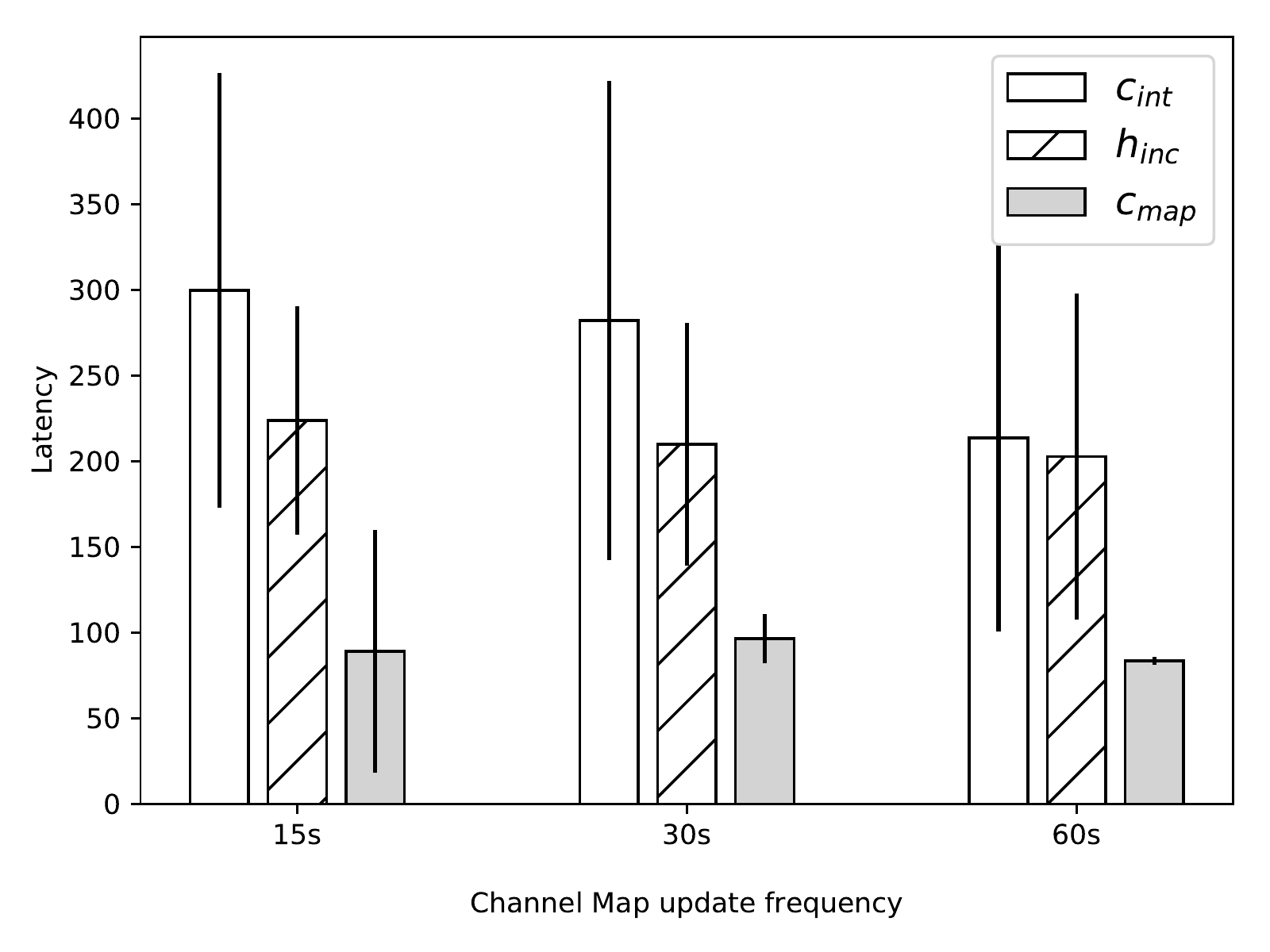}
\subcaption{}\label{lat_acc}
\end{minipage}\qquad
\begin{minipage}[b]{.3\textwidth}
\includegraphics[width=2in]{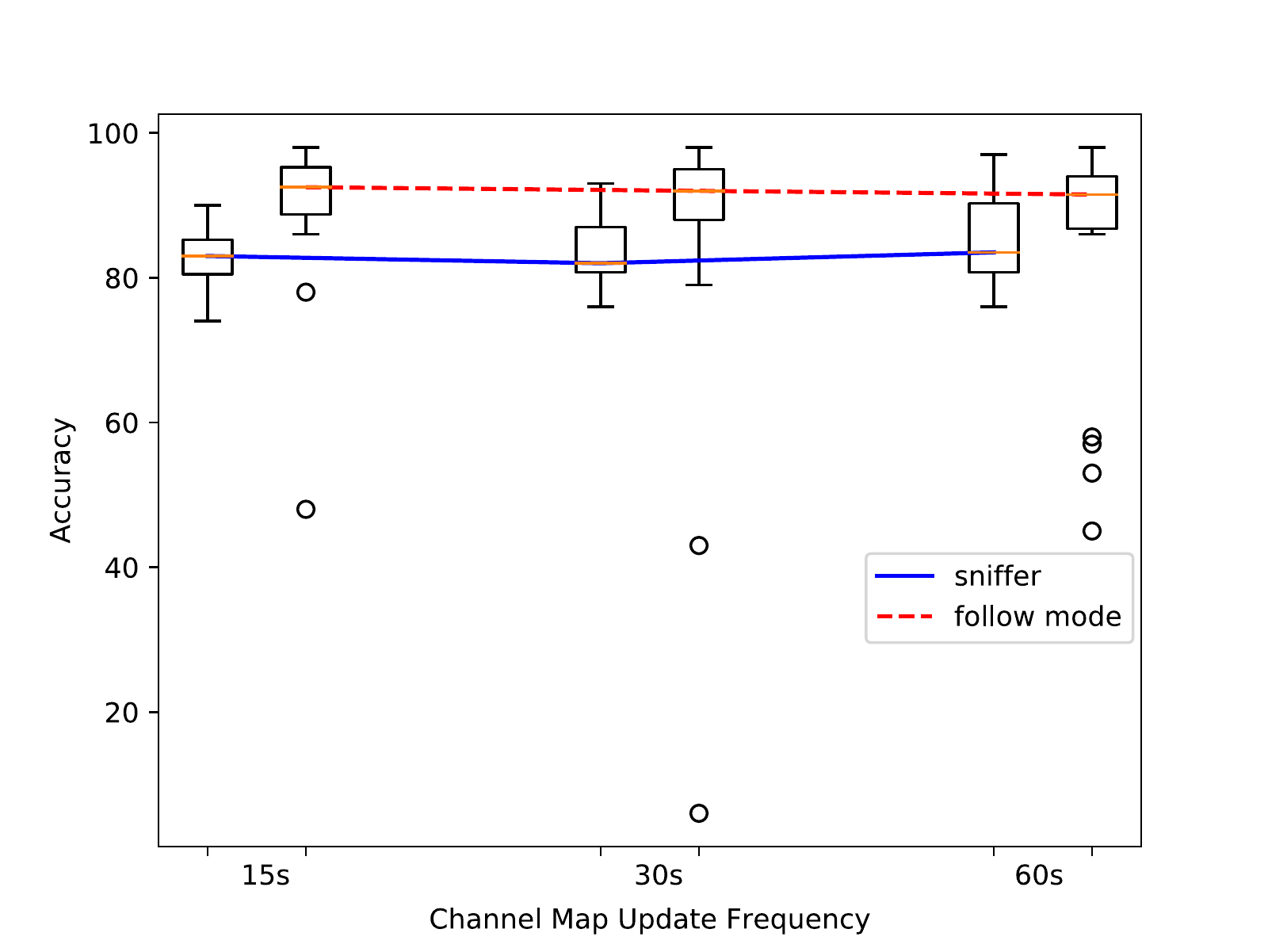}
\subcaption{}\label{100ms}
\end{minipage}\qquad
\begin{minipage}[b]{.3\textwidth}
\includegraphics[width=2in]{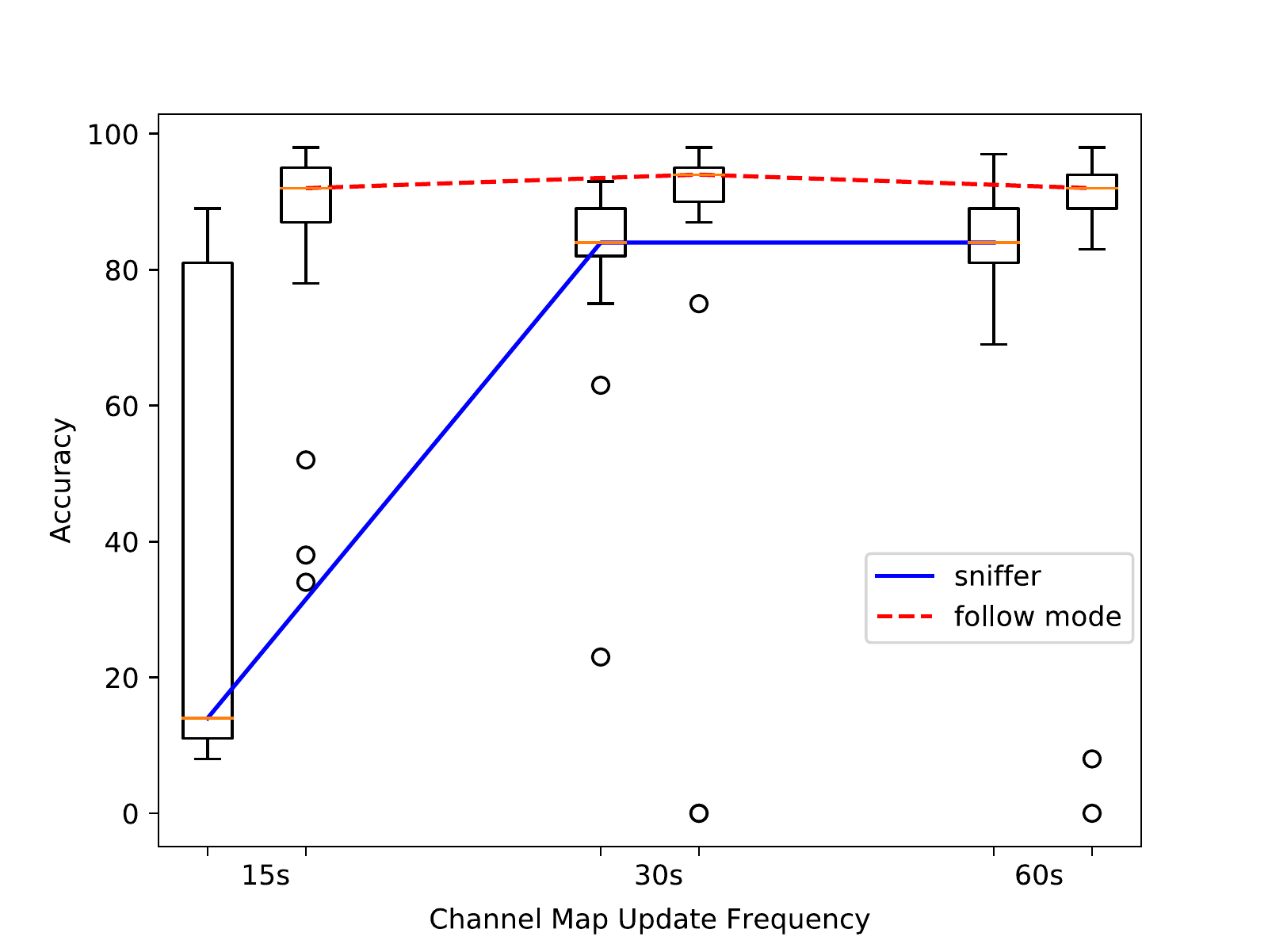}
\subcaption{}\label{250ms}
\end{minipage}
\caption{(a) latency in deriving AFH parameters; (b) sniffing accuracy in 100ms data sending rate; (c) sniffing accuracy in 250ms setting.}
\end{figure*}
\subsection{Performance Analysis Of AFH Parameter Calculation}
First, we present the theoretical analysis of computation overhead of determining each AFH parameter. For ${c_{int}}$, we need to observe the connection for at least ${2T}$ i.e., $2 \times 37$ hops. For accuracy, we observe the connection for ${nT}$ interval and compare the hopping pattern for $n$ times, where ${n > 2}$. In so doing, we increase the accuracy but incurs higher latency. In a non-ideal case where channel map updates during ${c_{int}}$ calculation, this will increase the time taken for determining ${c_{int}}$. 
For ${h_{inc}}$, ideally observing 3 channels in ${3T}$ shall yield the correct hop increment. However, if we want more accuracy, we observe ${k}$ different channels for ${kT}$ time, where ${k \geq 3}$. Next for ${c_{map}}$, we first pick a truly mapped channel and then continue to hop for $T$, which theoretically would give us ${c_{map}}$. However, if there is a channel map update during ${c_{map}}$ calculation, the hopping pattern will change and we will have wrong values in ${c_{map}}$. Thus, the latency for determining ${c_{map}}$ could be any multiplier of period $T$.

Experimentally, the accuracy of determining the AFH parameters is given in Table \ref{acc}. Besides, Fig. \ref{lat_acc} shows the latency (calculated in terms of hops) of correctly determining AFH parameters for same settings of ${c_{int}}$ and ${h_{inc}}$ but different frequencies of ${c_{map}}$ update. From Fig. \ref{lat_acc}, we see that the latency decreases as the frequency of ${c_{map}}$ update increases which justifies our previous reasoning.
\subsection{Performance Evaluation of Packet Sniffing}
We now evaluate the performance of our system in sniffing data packets and compare it with the Ubertooth One in follow mode. Two separate settings are examined where 100 data packets are transmitted every 100ms and every 250ms, respectively. During data transmissions, we update BLE devices' channel map based on a predefined frequency. From Fig. \ref{100ms} and Fig. \ref{250ms}, we observe that the accuracy for our sniffer increases as the frequency of ${c_{map}}$ update decreases. This is because the connection stays stable (i.e., AFH parameters remain unchanged) for long time and we do not need to run our algorithm frequently. On the other hand, the Ubertooth One follow mode exhibit the opposite trend. This is because the follow mode relies on BLE control packets to keep track of channel map.
\begin{table}
\begin{center}
\begin{tabular}{ |p{1.5cm}|p{1cm}|p{1cm}|p{1cm}|  }
 \hline
 Parameter & 15s   &30s   &60s\\
 \hline
  ${c_{int}}$     &100\%    &100\%   &100\%\\
  ${h_{inc}}$    &100\%     &96\%  &96\%\\
  ${c_{map}}$  &100\%     &96\%   &96\%\\
 \hline
\end{tabular}
\end{center}
 \caption{\label{acc}Accuracy for determining BLE AFH parameters}
 \end{table}

However, the performance of follow mode shows a larger variance than ours. The reason is that if the follow mode misses the packet containing ${c_{map}}$ update, it will not be able to re-sync until it receives the next ${c_{map}}$ update packet. On the contrary, our sniffer does not rely on BLE control packets. Whenever it gets de-synchronized with the BLE connection, it re-runs the ${c_{map}}$ algorithm to resume following the BLE connection.

\section{Conclusion}
In this work, we presented algorithms to determine the BLE connection parameters to crack its adaptive frequency hopping protocol, which leads to accurately sniffing BLE long-lived connections. We implemented our algorithms in the open source platform--Ubertooth One and evaluated its performance against a benchmark scheme. Our experimental results demonstrated the feasibility of the proposed algorithms in terms of low latency and high accuracy. We see that our algorithms can calculate AFH parameters with 96\% accuracy and sniff more than 80\% of the transmitted data.

\bibliography{gc_sopan_v3}
\bibliographystyle{IEEEtran}

\end{document}